\documentclass
[superscriptaddress,secnumarabic,amssymb,amsmath,nobibnotes,aps,prd,showkeys,nofootinbib,nopacsnumber,onecolumn,10pt]{revtex4}%
\usepackage{graphicx}
\usepackage{bm}
\usepackage{xcolor}
\usepackage{amsmath}
\usepackage{amsfonts}
\usepackage[english]{babel}
\usepackage{amssymb}
\usepackage[utf8]{inputenc}%
\providecommand{\U}[1]{\protect\rule{.1in}{.1in}}

\newcommand{\be}{\begin{equation}}
\newcommand{\ee}{\end{equation}}

\newcommand{\mincir}{\raise
-3.truept\hbox{\rlap{\hbox{$\sim$}}\raise4.truept\hbox{$<$}\ }}
\newcommand{\magcir}{\raise
-3.truept\hbox{\rlap{\hbox{$\sim$}}\raise4.truept\hbox{$>$}\ }}

\begin{document}
\title{Data-Driven Conservation Laws for Newtonian Systems: Physics-Informed Neural
Networks and Eisenhart-Duval Geometry}
\author{Andronikos Paliathanasis}
\email{anpaliat@phys.uoa.gr}
\affiliation{Institute of Systems Science, Durban University of Technology, Durban 4000,
South Africa}
\affiliation{Centre for Space Research, North-West University, Potchefstroom 2520, South Africa}
\affiliation{Centro de Investigaci\'on, Innovaci\'on y Creaci\'on (CIIC), Universidad
Cat\'olica de Temuco, Temuco, Chile}
\affiliation{Departamento de Ciencias Matem\'{a}ticas y F\'{\i}sicas, Facultad de
Ingenier\'{\i}a, Universidad Cat\'olica de Temuco, Temuco, Chile}
\affiliation{National Institute for Theoretical and Computational Sciences (NITheCS), South Africa}

\begin{abstract}
We introduce a Physics-Informed Neural Network approach for the data-driven
reconstruction of the symmetries and conservation laws of Newtonian dynamical
systems, based on the geometrization of the dynamics through the
Eisenhart-Duval lift. Within the Eisenhart-Duval lift, the symmetry conditions
reduce to the linear Killing or Conformal Killing equations of the extended
metric. We combine a Hamiltonian neural network, which reconstructs the
potential function from the data, with a random-feature neural network, which
solves the symmetry conditions for the collineations of the extended geometry.
From the learned vector fields we construct the Lie algebra and, through
Noether's theorem, the corresponding conserved quantities, while the
derivation of the curvature invariants for the extended geometry provide allow
us to perform an independent consistency test. We apply the method to the
one-dimensional Ermakov-Pinney equation and to the one-dimensional harmonic
oscillator, with and without a quartic perturbation, by using different level
of noise in the phase-space data. For the Ermakov-Pinney equation, the three
Killing vectors of the flat extended plane, and the corresponding nonlocal
conservation laws, are recovered for, while for the harmonic oscillator, the
ten conformal Killing vectors which form the $so(3,2)$ Lie algebra are
determined for noise levels up to $10\%$. When the quartic term is added, the
broken symmetries are identified as approximate symmetries.

\end{abstract}
\date{\today}
\maketitle

\section{Introduction}

\label{sec1}

Newton's second law expresses the dynamics of physical systems as second-order
ordinary differential equations. The solutions of these equations describe the
motion in a wide range of physical systems, from celestial mechanics to
molecular dynamics, and allow us to make predictions about their evolution.
The simplest systems in Newtonian mechanics are the free particle, described
by the equation $\ddot{q}=0$, and the harmonic oscillator, whose equation of
motion $\ddot{q}+\omega^{2}q=0$ describes small oscillations around a stable
equilibrium. Here $q=q\left(  t\right)  $ is the dependent variable which
describes the position of the particle, $t$ is the independent variable that
denotes the time parameter, $\omega^{2}$ is the frequency of the oscillator
and a dot represents the derivative with respect to the independent variable.

The solution of the free particle is a linear function of time, while the
solution of the oscillator is expressed in terms of the trigonometric
functions, which are defined by infinite power series. Despite the different
form of their solutions, the two systems are equivalent, that is, they are
related by a point transformation, and both admit the maximal number of Lie
point symmetries \cite{mah1,mah2,mah3}. This equivalence is revealed by the
symmetries of the equations of motion, and it is a first indication that the
symmetry structure of a dynamical system carries information which is not
evident from the equations themselves.

A systematic method for the study of differential equations is the symmetry
analysis introduced by Sophus Lie \cite{lie1,lie2,lie3}. A Lie symmetry vector
is the generator of a one-parameter group of transformations which leaves the
differential equation invariant, that is, it maps solutions into solutions.
The main novelty of the Lie symmetry analysis is that the invariants of the
symmetry vectors can be used to simplify the differential equation, by
reducing the order of an ordinary differential equation, or the number of
independent variables for the case of partial differential equations
\cite{olver,kumei,ibra,bk4}. Therefore, the solution of the dynamical system
can be expressed in algebraic form or by quadratures, or the system can be
transformed into the form of another well-known system, as in the relation
between the oscillator and the free particle discussed before. Thus,
symmetries have been used to classify dynamical systems according to the
admitted group properties \cite{Ovsi}.

For dynamical systems which follow from a variational principle, the Lie
symmetries which also leave the action integral invariant, up to a boundary
term, are characterized as Noether symmetries\ \cite{noe} and they are the
origin of conserved quantities, the Noether charges. For a recent review on
Noether's theorems we refer the reader to \cite{ren1}. Conservation laws are
fundamental in the study of dynamical systems since they define the surfaces
within the phase space on which the motion takes place, they permit the
reduction of the dynamical system, and they reveal information about its
integrability properties. Since conservation laws follow from symmetries
through Noether's theorem, symmetry analysis is an essential tool for the
study of nonlinear differential equations, which is why it has been widely
used for the study of nonlinear phenomena, in physical science and beyond, see
for instance
\cite{app1,app2,app3,app4,ap11,ap12,app5,app6,app7,app8,app9,app10,ap16,ap14,ap15}
and references therein.

The determination of Lie and Noether symmetries requires the solution of a
system of partial differential equations \cite{Ovsi}, the symmetry conditions,
which depend explicitly on the form of the equations of motion. For Newtonian
dynamical systems these conditions have a geometric interpretation. Indeed,
the symmetry vectors of the equations of motion of a particle moving in a
Riemannian space are related to the collineations of the kinetic metric, while
the potential function introduces additional constraints \cite{tsa1}.\ On the
other hand, the Noether symmetries for the Newtonian systems with a
variational principle are related to the admitted elements of the homothetic
algebra of the flat space where the motion occurs \cite{tsa2}, or to the
elements of the conformal algebra for the background geometry in the case of
constrained systems with zero value for the energy \cite{tsa3}. These
geometric characteristics become clear in the framework of the Eisenhart-Duval
lift \cite{el1,duval}, where the trajectories of the dynamical system are
described by the equivalent set of geodesic equations for an extended
manifold, where the potential component has been absorbed in the new extended
metric tensor. Therefore, the Lie and Noether symmetries are directly related
to the collineations of the extended geometry without any additional
constraint. The collineations can be used for the construction of conservation
quantities for the geodesic system \cite{katz,am1,am2}.

In the Eisenhart-Duval approach, the geometrization is achieved by introducing
additional degrees of freedom \cite{el1}, which absorb the potential function
of the dynamical system. The resulting equivalent system is described by the
geodesic equations of a higher-dimensional Riemannian or Lorentzian manifold.
In order to recover the original system, the extended geometry must admit
Killing vector fields along the new coordinates. The corresponding
conservation laws constrain the evolution of the new degrees of freedom, which
is necessary in order to map the geodesics to the solutions of the original
system. In \cite{el1}, Eisenhart introduced the method by adding one new
degree of freedom, which leads to the Riemannian lift. Later, in \cite{duval},
a lift was formulated with two additional degrees of freedom, such that the
extended geometry is a Lorentzian spacetime; this is known as the Lorentzian
lift. Nevertheless, the lift is not unique, and other formulations can lead to
new results \cite{annew}. The Eisenhart-Duval lift has been applied to the
determination of conservation laws in a wide range of dynamical systems, and
it has been shown that it reveals hidden symmetries and conservation laws of
the original system \cite{anhidden}. Furthermore, the Eisenhart-Duval lift
provides a different direction for the geometric linearization of dynamical
systems which fail to satisfy the Lie symmetry conditions for linearization.
Indeed, by using the Eisenhart-Duval lift, it was shown in \cite{common1} that
a large class of gravitational models share a common solution space, that of
the free particle in a flat extended geometry, such that each physical system
corresponds to a different representation of the common solution. For other
applications we refer the reader to \cite{ed1,ed2,ed3,ed4,ed6,ed7}.

Machine learning techniques have become an important tool for the study of
dynamical systems from the data. Physics-Informed Neural Networks \cite{pinn0}
are of special interest since they force the network to satisfy the given
physical laws of motion \cite{pinn1,pinn2,pinn3,pinn4,pinn5,pinn6,pinn7,pinn8}%
.\ A first application of PINNs is to Hamiltonian mechanics, introduced in
\cite{hnn1}, where Hamilton's equations were enforced to reconstruct the
Hamiltonian function, and consequently, the conservation law of energy for
dynamical systems directly from the data. The Lagrangian formalism within
a\ PINN was introduced later in \cite{hnn2}. Moreover, in \cite{hnn3},
conserved quantities have been determined from trajectories without the
knowledge of the equations of motion. As far as the admitted symmetries are
concerned, there are various approaches for the reconstruction of symmetries
directly from the data \cite{hnn4,hnn5}.

The reconstruction of the symmetries and of the conservation laws is performed
through the introduction of the residual within the loss function. In this
work we introduce a new framework, by employing the Eisenhart-Duval framework
for the derivation of symmetries and conservation laws of Newtonian dynamical
systems. Within the Eisenhart-Duval lift, the derivation of the symmetries is
equivalent to the derivation of the collineations for the extended
geometry.\ This is a linear problem, and can be approached with a linear model
such as the random-featured expansion described in \cite{hnn6}. The structure
of the paper is as follows.

In Section \ref{sec2} we present the basic properties and elements of the
Eisenhart-Duval lift, for the geometric description of dynamical systems.
Moreover, we introduce the symmetry structure of Riemannian and Lorentzian
manifolds, and specifically the Killing symmetries, the Homothetic symmetries
and the Conformal Killing symmetries, which are related through Noether's
theorem to conservation laws for the massive and the massless (null) geodesic
Lagrangians. We review previous results and we demonstrate the application of
the Eisenhart-Duval lift for the derivation of local and hidden conservation
laws. Applications to Newtonian Mechanics are presented. In Section \ref{sec3}
we introduce the main approach which we employ for the reconstruction of
conservation laws for dynamical systems. In particular, we introduce a
physics-informed neural network that learns the potential function of a
Newtonian Hamiltonian system directly from trajectory data. The learned
potential is used to construct the corresponding Eisenhart-Duval metric, in
which the dynamical system is described by the geodesic equations. We examine
the geometric structure of the extended metric by calculating curvature
invariants and tensors, namely the Ricci scalar, the Cotton-York tensor and
the Weyl tensor, which reveal important information about the characteristics
of the extended geometry. We then introduce a random-feature network that
determines the symmetry vector fields from a linear null-space problem, along
with a bootstrap criterion that selects the number of symmetries, and we
extract the Lie algebra and the corresponding conserved quantities by applying
Noether's theorem for the geodesic Lagrangian. In Section \ref{sec4} we
demonstrate the application of the networks for the data-driven determination
of conservation laws, using mock data. Specifically, we consider noisy
trajectories for the one-dimensional Ermakov-Pinney equation and the
one-dimensional harmonic oscillator, two models which lead to maximally
symmetric geometries after the application of the Einsenhart-Duval lift.
Finally, in Section \ref{sec5} we summarize our results and we draw our conclusions.

\section{Preliminaries}

\label{sec2}

In this section we briefly discuss the main geometric framework used in this
study. Specifically we introduce the collineations which provides conservation
laws for the geodesic equations, and the Eisenhart-Duval lift for the
geometrization of Hamiltonian systems.

\subsection{Symmetries of the Metric and of the Geodesic Equations}

We briefly introduce the symmetries for the metric tensor and how they can be
used to construct conservation laws for the geodesic equations.

\subsubsection{Symmetries of the Metric}

Consider a Riemannian manifold endowed with the metric tensor $g_{\mu\nu
}\left(  x^{\mu}\right)  $, then we shall say that the generator $\xi^{\mu
}=\xi^{\mu}\left(  x^{\kappa}\right)  $ of the infinitesimal one parameter
point transformation $x^{\prime\mu}=x^{\mu}+\varepsilon\xi^{\mu}\left(
x^{\kappa}\right)  $ leaves invariant the length, distances, and angles of
geometric objects if and only if \cite{bar3}
\begin{equation}
\mathcal{L}_{\xi}g_{\mu\nu}=0\text{.} \label{sc.01}%
\end{equation}
where $\mathcal{L}_{\xi}$ denotes the Lie derivative with respect to the
vector field $\xi^{\mu}$, which for a second-rank tensor field reads%
\[
\mathcal{L}_{\xi}g_{\mu\nu}=g_{\mu\nu;\kappa}\xi^{\kappa}+g_{\mu\kappa}%
\xi_{~,\nu}^{\kappa}+g_{\kappa\nu}\xi_{~,\mu}^{\kappa}\,\text{.}%
\]
If condition (\ref{sc.01}) is true, $\xi^{\mu}$ acts as a symmetry of the
metric tensor and it is characterized as a Killing vector (KV). For the two
dimensional Euclidian space, the Killing symmetries form the $e\left(
2\right)  $ Lie algebra, also known as Bianchi VII$_{0}$ algebra, while for
the three-dimensional Euclidian space the Killing vectors form the $e\left(
3\right)  $ Lie algebra. The four-dimensional Minkowski geometry is invariant
under the ten elements of the Poincar\'{e} group. Killing symmetries
characterize the geometry and reveal important features when they exist.

The admitted number of Killing symmetries in Riemannian spaces is not
arbitrary; two-dimensional geometries can admit zero, one or three Killing
vector fields, with the latter being true for the maximally symmetric spaces.
These are the $e\left(  2\right)  $ for the flat space, as we discussed
before, and the $so\left(  3\right)  ,so\left(  2,1\right)  $ for the space of
constant curvature. In general the $n$-dimensional maximally symmetric spaces
possess $\frac{1}{2}n(n+1)$ Killing symmetries, which form the $e\left(
n\right)  $ or $so\left(  n+1\right)  /so\left(  n,1\right)  $ Lie algebras respectively.

Conformal killing vectors (CKVs) are the generators of infinitesimal one
parameter point transformations which do not preserve the length and the
distances, but they preserve the angles. They are defined as \cite{bar3}
\begin{equation}
\mathcal{L}_{\xi}g_{\mu\nu}=2\psi\left(  x^{\kappa}\right)  g_{\mu\nu}%
,~\psi\left(  x^{\kappa}\right)  =\frac{1}{n}\xi_{;\mu}^{\mu}. \label{sc.02}%
\end{equation}
We remark that when $\psi\left(  x^{\kappa}\right)  =0$, the Killing equation
(\ref{sc.01}) is recovered, which means that KVs are particular cases of the
CKVs. In general if $\psi\left(  x^{\kappa}\right)  \neq0$, the field
$\xi^{\mu}$ is called a proper CKV, while in the case where $\psi\left(
x^{\kappa}\right)  $ is a constant, the $\ $field $\xi^{\mu}$ is characterized
as a Homothetic killing vector (HKV).

The admitted algebra for the CKVs characterizes the Riemannian manifold. As
with the case of KVs, the maximum admitted algebra of CKVs has dimension
$\frac{1}{2}\left(  n+1\right)  \left(  n+2\right)  $, for $n>2$, the
$so\left(  n+1,1\right)  ,$ admitted by conformally flat spaces. Therefore,
the three-dimensional flat space admits ten CKVs, which form the $so\left(
4,1\right)  $ Lie algebra.\ On the other hand, two-dimensional spaces admit an
infinite number of CKVs, while the one-dimensional space does not admit any
proper CKV.

A Riemannian geometry is conformally flat if and only if, for $n>3$, it has
zero Weyl tensor \cite{ky}%
\begin{equation}
C_{\mu\nu\kappa\lambda}=R_{\mu\nu\kappa\lambda}-\frac{2}{n-2}\left(
g_{\mu\lbrack\kappa}R_{\lambda]\nu}-g_{\nu\lbrack\kappa}R_{\lambda]\mu
}\right)  +\frac{2}{\left(  n-1\right)  \left(  n-2\right)  }Rg_{\mu
\lbrack\kappa}g_{\lambda]\nu},
\end{equation}
while for $n=3$, the components of the Cotton-York tensor \cite{ky}
\begin{equation}
C_{\mu\nu\kappa}=R_{\nu\mu;\kappa}-R_{\nu\kappa;\mu}+\frac{1}{4}\left(
R_{;\mu}g_{\nu\kappa}-R_{;\kappa}g_{\nu\mu}\right)  ,~n=3,
\end{equation}
vanish. Finally, all two-dimensional Riemannian manifolds are always
conformally flat.

\subsubsection{Conservation laws for the Geodesic Equations}

It is well known that there exists a unique relation between the symmetries of
the metric and the symmetries of the geodesic equations and their
corresponding conservation laws \cite{katz,am1,am2}. For instance the
conservation law of momentum for the free particle follows from the
translation symmetry of the Euclidean spacetime. The group of translations and
the group of rotations form the group of isometries or Killing vectors of the
Euclidean space, while the conservation law for the angular momentum is
related to the rotation symmetries of the flat space.

From Noether's theorem, it follows that the quantity\footnote{It is imporant
to mention that in this study we focus on conservation laws which are
independent of time, $I_{,t}=0$, and are linear in the momentum. More general
conservation laws, polynomial in the momenta, can be constructed for the
geodesic equations by means of Killing tensors of higher rank.} $I=p_{\mu}%
\xi^{\mu}$ is a conservation law for the geodesic equations, if $\xi^{\mu}$ is
a KV, that is, $\frac{dI}{dt}=I_{,t}+\left\{  I,\mathcal{H}_{G}\right\}  =0,$
hence $\left\{  I,\mathcal{H}_{G}\right\}  =0$, where $\mathcal{H}_{G}$ is the
Hamiltonian function for the geodesic equations $\mathcal{H}_{G}=\frac{1}%
{2}g^{\mu\nu}p_{\mu}p_{\nu}$.

Condition $\frac{dI}{dt}=0$, is equivalent to the Killing equation
(\ref{sc.01}). Nevertheless, the geodesic Hamiltonian is not invariant under
conformal transformations except if the \textquotedblleft
Energy\textquotedblright\ is zero, that is, when it describes massless (null)
geodesics. In this case, under a conformal transformation, the volume
continues to be zero under conformal transformations and consequently
conservation laws can be constructed. Hence, if $\xi^{\mu}$ is a CKV, the
symmetry condition is relaxed as $\frac{dI}{dt}=2\psi\left(  x^{\kappa
}\right)  \mathcal{H}_{G}$, $\operatorname{mod}\mathcal{H}_{G}=0$, where
$\psi\left(  x^{\kappa}\right)  $ is the conformal factor. Therefore, we end
with the condition $\left\{  I,\mathcal{H}\right\}  =2\psi\left(  x^{\kappa
}\right)  \mathcal{H}$, which is equivalent \ to the conformal condition
(\ref{sc.02}).

\subsection{Eisenhart-Duval lift}

Consider an $n$-dimensional dynamical system described by the Hamiltonian
function%
\begin{equation}
\mathcal{H}=\frac{1}{2}g^{\mu\nu}\left(  q^{\kappa}\right)  p_{\mu}p_{\nu
}+V\left(  q^{\kappa}\right)  =E, \label{ns.00}%
\end{equation}
where $q^{\mu},~p_{\mu}$ are the position and the momentum terms respectively,
$V\left(  q^{\mu}\right)  $ is the potential function that describes the
dynamics and $g^{\mu\nu}\left(  q^{\kappa}\right)  $ is a second-rank tensor
that defines the kinetic term. In Newtonian physics, $g^{\mu\nu}\left(
q^{\kappa}\right)  $ represents the $n$-dimensional Euclidian flat space, and
$E$ is the value for the energy. The equations of motion read
\begin{align}
\dot{q}^{\mu}  &  =g^{\mu\nu}p_{\mu},\label{ns.01}\\
\dot{p}_{\mu}  &  =F_{\mu},~ \label{ns.02}%
\end{align}
in which $F_{\mu}=-V_{,\mu}$, is the force term.

We introduce the Eisenhart-Duval lift as a geometric framework for
representing dynamical systems, with particular emphasis on Newtonian
dynamics. Within this approach, the study of the dynamical system is
reformulated in terms of the geometric properties of an extended space. The
collineations of the extended metric can be used to construct local and
nonlocal conservation laws for the original system \cite{anhidden}, while
properties such as the linearizability of the dynamical system can be
investigated by examining the geometric invariants of the lifted space
\cite{common1}.

\subsubsection{Riemannian lift}

We proceed with the Riemannian lift \cite{el1} and we consider the $\left(
n+1\right)  $-dimensional dynamical system with Hamiltonian function
\begin{equation}
\mathcal{H}_{R}=\frac{1}{2}g^{\mu\nu}p_{\mu}p_{\nu}+\frac{1}{2}V\left(
q^{\kappa}\right)  p_{z}^{2}=E_{R}. \label{ns.04}%
\end{equation}
The latter function describes the geodesic equations for the $\left(
n+1\right)  $-dimensional Riemannian manifold with line element%
\begin{equation}
ds_{R}^{2}=g_{\mu\nu}\left(  q^{\kappa}\right)  dq^{\mu}dq^{\nu}+\frac
{1}{V\left(  q^{\kappa}\right)  }dz^{2}\,, \label{ns.05}%
\end{equation}
and geodesic equations%
\begin{align}
\dot{q}^{\mu}  &  =g^{\mu\nu}p_{\nu},\label{nn.06}\\
\dot{z}  &  =V\left(  q^{\kappa}\right)  p_{z},\label{nn.07}\\
\dot{p}_{\mu}  &  =\frac{1}{2}F_{\mu}p_{z}^{2},\label{nn.08}\\
\dot{p}_{z}  &  =0. \label{nn.09}%
\end{align}
The momentum term $p_{z}$ is a conservation law for the geodesic equations,
related to the Killing symmetry $\partial_{z}$ for the extended metric
(\ref{ns.05}). Therefore, the solution trajectories of the geodesic equations
(\ref{nn.06})-(\ref{nn.09}) are solutions for the original dynamical system
(\ref{ns.01}), (\ref{ns.02}), when $E_{R}=E$ and $\frac{1}{2}p_{z}^{2}=1$ .
Due to the Killing symmetry $\partial_{z}$, the extended geometry is foliated
and the extended degrees of freedom $z,$ $p_{z}$, play no role in the
evolution of the original system.

\subsubsection{Lorentzian lift}

On the other hand, in the Lorentzian lift approach \cite{duval}, we consider
the $\left(  n+2\right)  $-dimensional dynamical system with Hamiltonian
function%
\begin{equation}
\mathcal{H}_{L}\equiv\frac{1}{2}g^{\mu\nu}p_{\mu}p_{\nu}+V\left(  q^{\kappa
}\right)  p_{v}^{2}+p_{u}p_{v}=E_{L}, \label{nn.10}%
\end{equation}
and equations of motion%
\begin{align}
\dot{q}^{\mu}  &  =g^{\mu\nu}p_{\mu},\\
\dot{u}  &  =2V\left(  q^{k}\right)  p_{u}+p_{v},\\
\dot{v}  &  =p_{u},\\
\dot{p}_{\mu}  &  =F_{\mu}p_{u}^{2},\\
\dot{p}_{u}  &  =0,\\
\dot{p}_{v}  &  =0,
\end{align}
where $p_{u}$ and $p_{v}$ are conserved quantities. We remark that the
original system (\ref{ns.01}), (\ref{ns.02}) is recovered when $p_{u}^{2}%
=1~$and $p_{u}p_{v}-E_{L}=E$. Similarly to before, the extended degrees of
freedom $u,v,p_{u}$ and $p_{v}$ play no role in the dynamic evolution for the
rest of the variables.

Function (\ref{nn.10}) is the geodesic Hamiltonian for the $\left(
n+2\right)  $-dimensional Riemannian geometry of pp-wave like form%
\begin{equation}
ds_{L}^{2}=g_{\mu\nu}\left(  q^{\kappa}\right)  dq^{\mu}dq^{\nu}%
+2dudv-2V\left(  q^{\kappa}\right)  du^{2}. \label{nn.11}%
\end{equation}
Without loss of generality in the Lorentzian lift we can assume $E_{L}=0$,
such that the Hamiltonian function (\ref{nn.10}) describes massless (null)
geodesic equations. This choice is particularly useful because the null
geodesic equations are invariant under conformal transformations, up to a
reparametrization of the geodesics. Consequently, conformal symmetries of the
lifted geometry can be used to construct conservation laws for the null
geodesic dynamics. The non-uniqueness of the lifting approach is discussed in
detail in \cite{annew}, where different families of extended geometries are
introduced, leading to new results.

\subsection{Newtonian systems}

We continue with the application of the Eisenhart-lift to two systems of
special interest, the Ermakov-Pinney equation, and the harmonic oscillator.
These two systems have the property of being linearizable within the
Einsenhart-Duval framework.

\subsubsection{Ermakov-Pinney Equation}

The one-dimensional Ermakov-Pinney equation is described by the Hamiltonian
function
\begin{equation}
\mathcal{H}_{EP}=\frac{1}{2}p^{2}+\frac{1}{2}\frac{\mu}{q^{2}}=E_{EP},
\label{ep.01}%
\end{equation}
with equations of motion\qquad\
\begin{align}
\dot{q}  &  =p,\\
\dot{p}  &  =\frac{\mu}{q^{3}}.
\end{align}
with solution $q=\sqrt{q_{1}^{2}\left(  t\right)  +q_{2}^{2}\left(  t\right)
},$ where $q_{1,2}\left(  t\right)  $ are the two solutions of the free
particle $\ddot{q}_{1,2}=0$.

We employ the Riemannian lift, and we define the equivalent geodesic
Hamiltonian%
\begin{equation}
\mathcal{H}_{EP,R}=\frac{1}{2}p^{2}+\frac{1}{2}\left(  \frac{2q^{2}}{\mu
}\right)  p_{z}^{2}=E_{EP,R},
\end{equation}
which recovers the original system when $p_{z}=1$ and $E_{EP}=E_{EP,R}$. The
kinetic metric for the geodesic Lagrangian reads
\begin{equation}
ds^{2}=dq^{2}+\left(  \frac{2q^{2}}{\mu}\right)  dz^{2},
\end{equation}
where we calculate the Ricci scalar $R=0$.\ From this we conclude that the
space is the flat space and admits three KVs, two translations and one
rotation. In Cartesian coordinates $X=q\cos z,~Y=q\sin z,$ the vector fields
read $\left\{  \partial_{X},\partial_{Y},Y\partial_{X}-X\partial_{Y}\right\}
$ with conservation laws\footnote{Since we are interested in time-independent
conservation laws, we omit the presentation of the symmetries generated by the
HKV, which generate the admitted$\ sl(2,R)$ Lie algebra.}%
\begin{align}
I_{1}  &  =Yp_{X}-Xp_{Y},\\
I_{2}  &  =p_{X},\\
I_{3}  &  =p_{Y}.
\end{align}
The transformation from the polar to Cartesian coordinates is a nonlocal
transformation, which transforms the equations for the free particle $\ddot
{X}=0$, into the Ermakov-Pinney equation. The inverse transformation
$q=\sqrt{X^{2}+Y^{2}}$ defines the solution.

We show that the Ermakov-Pinney equation can be linearized through the
extended geometry.

\subsubsection{Harmonic Oscillator}

Consider now the harmonic oscillator with Hamiltonian function
\begin{equation}
\mathcal{H}_{HO}=\frac{1}{2}p^{2}+\frac{1}{2}\omega^{2}q^{2}=E_{HO}%
\end{equation}
and equations of motion%
\begin{align}
\dot{q}  &  =p,\\
\dot{p}  &  =-\omega^{2}q.
\end{align}
We employ the Lorentzian lift and we introduce the (null) geodesic equivalent
Hamiltonian%
\begin{equation}
\mathcal{H}_{HO,L}\equiv\frac{1}{2}p^{2}+\frac{1}{2}\omega^{2}q^{2}p_{v}%
^{2}+p_{u}p_{v}=0.
\end{equation}
For the three-dimensional extended geometry we calculate the Cotton-York
tensor and we find that it has zero components, consequently, the extended
space is conformally flat. The extended space possesses a ten dimensional
Conformal algebra, the $so\left(  3,2\right)  $, with elements%
\begin{align}
K^{1}  &  =\cos\left(  \omega u\right)  \partial_{q}+q\sin\left(  \omega
u\right)  \partial_{v}\\
K^{2}  &  =\sin\left(  \omega u\right)  \partial_{q}-q\cos\left(  \omega
u\right)  \partial_{v},\\
K^{3}  &  =\partial_{u},\\
K^{4}  &  =\partial_{v}%
\end{align}
which form the four-dimensional Killing algebra, and
\begin{align}
CKV^{1}  &  =q\partial_{q}+2v\partial_{u},\\
CKV^{2}  &  =qv\partial_{q}-\frac{q^{2}}{2}\partial_{u}+\left(  v^{2}%
-\omega^{2}\frac{q^{4}}{4}\right)  \partial_{v},\\
CKV^{3}  &  =\left(  \cos\left(  \omega u\right)  +\frac{\omega q^{2}}{2}%
\sin\left(  \omega u\right)  \right)  \partial_{q}-q\cos\left(  \omega
u\right)  \partial_{v}+\frac{q}{2}\left(  2\omega v\sin\left(  \omega
u\right)  -\cos\left(  \omega u\right)  \omega^{2}q^{2}\right)  \partial
_{v},\\
CKV^{4}  &  =\left(  \sin\left(  \omega u\right)  -\frac{\omega q^{2}}{2}%
\cos\left(  \omega u\right)  \right)  \partial_{q}+q\sin\left(  \omega
u\right)  \partial_{v}-\frac{q}{2}\left(  2\omega v\cos\left(  \omega
u\right)  +\sin\left(  \omega u\right)  \omega^{2}q^{2}\right)  \partial
_{v},\\
CKV^{5}  &  =q\omega\sin\left(  2\omega u\right)  \partial_{q}+\cos\left(
2\omega u\right)  \partial_{u}+\omega^{2}q^{2}\cos\left(  2\omega u\right)
\partial_{v},\\
CKV^{6}  &  =q\omega\cos\left(  2\omega u\right)  \partial_{q}+\sin\left(
2\omega u\right)  \partial_{u}+\omega^{2}q^{2}\sin\left(  2\omega u\right)
\partial_{v}.
\end{align}

At this point it is important to mention that for the $n$-dimensional
isotropic harmonic oscillator the Lorentzian lift leads always to a $\left(
n+2\right)  $-dimensional geometry which is always conformally flat.

\section{Data-Driven Conservation Laws via Eisenhart--Duval Geometry}

\label{sec3}

\begin{figure}[ptbh]
\centering\includegraphics[width=1\textwidth]{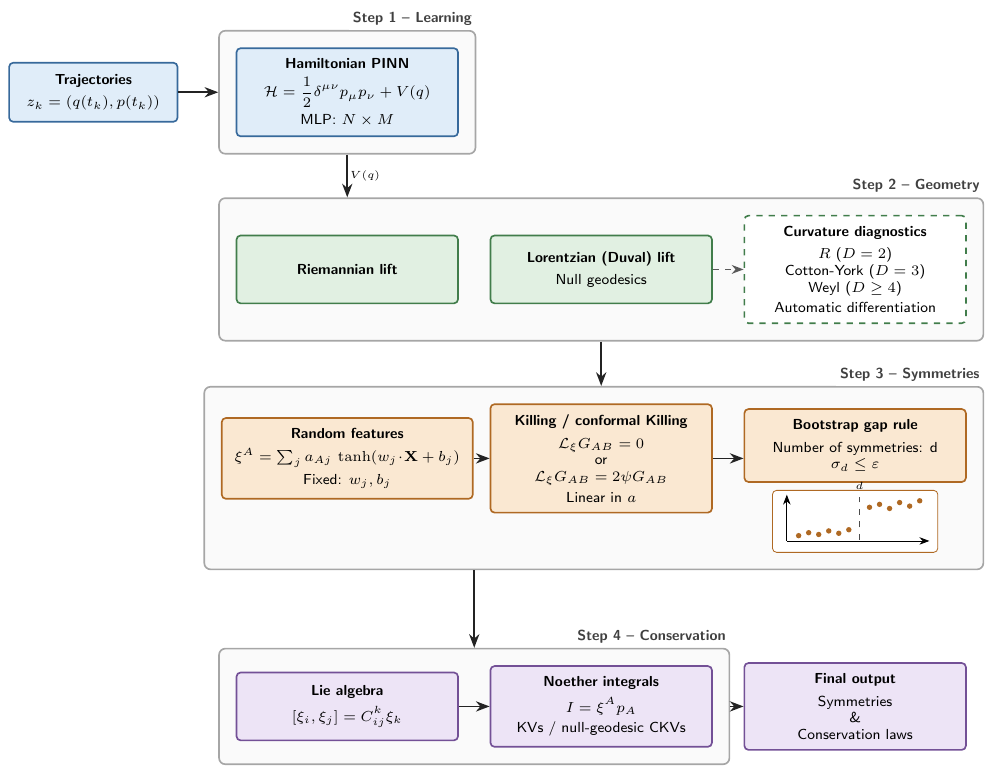}\caption{Flow
diagram for the data-drive reconstruction of the conservation laws and
symmetries for Newtonian systems via the Eisenhart-Duval geometry.}%
\label{fig0}%
\end{figure}

In this section we introduce the method that we employ for the determination
of the number of symmetries and the reconstruction of conservation laws for
dynamical systems directly from data. This framework consists of four steps,
(i) a PINN for the reconstruction of the Hamiltonian function; (ii) the
application of the Eisenhart-Duval lift and the reconstruction of the extended
geometry; (iii) a random-feature neural network that provides the KVs and the
CKVs for the extended metric, and as a final step (iv) the construction of the
conserved quantities and the determination of the commutators. Fig. \ref{fig0}
shows the flow diagram for the framework.

\subsection{Physics-Informed Neural Network}

We make the assumption that the observed data are points on the phase space
$z_{k}=\left(  q\left(  t_{k}\right)  ,p\left(  t_{k}\right)  \right)  $ for a
Newtonian system with Hamiltonian function (\ref{ns.00}) and $g_{\mu\nu}$
describes the flat space in Cartesian coordinates\footnote{In this framwork we
consider the simplest Hamiltonian, but the approach can also be extended to
more general systems.}, that is, $g_{\mu\nu}=\delta_{\mu\nu}$.

The potential function $V\left(  q\right)  $ is trainable and is learned using
a Hamiltonian Neural Network approach. The equations of motion (\ref{ns.01}),
(\ref{ns.02}) are integrated for a step number $K$, by using the
St\"{o}rmer-Verlet approach \cite{sv1,sv2}, that is, we integrate as follows
\begin{align}
q_{k+1}  &  =q_{k}+p_{k+\frac{1}{2}}\Delta t,\\
p_{k+\frac{1}{2}}  &  =p_{k}+\frac{\Delta t}{2}F\left(  q_{k}\right)  ,\\
p_{k+1}  &  =p_{k+\frac{1}{2}}+\frac{\Delta t}{2}F\left(  q_{k+1}\right)  ,
\end{align}
where $\Delta t$ is the step of integration.

We consider a dimensionless loss function, defined by the average difference
between the trajectories predicted by the learned Hamiltonian and the data,
i.e.
\begin{equation}
\mathcal{L}=\sum_{\mu}\frac{\left\langle \left(  \hat{q}^{\mu}\left(
t_{k+K}\right)  -q^{\mu}\left(  t_{k+K}\right)  \right)  ^{2}\right\rangle
}{\mathrm{Var}\left(  q^{\mu}\right)  }+\sum_{\mu}\frac{\left\langle \left(
\hat{p}_{\mu}\left(  t_{k+K}\right)  -p_{\mu}\left(  t_{k+K}\right)  \right)
^{2}\right\rangle }{\mathrm{Var}\left(  p_{\mu}\right)  },
\end{equation}
where $\left\{  \hat{q}^{\mu}\left(  t_{k+K}\right)  ,\hat{p}_{\mu}\left(
t_{k+K}\right)  \right\}  $ are the predictions of the network starting from
the data at $t_{k}$ and $\mathrm{Var}\left(  q^{\mu}\right)  ,~\mathrm{Var}%
\left(  p^{\mu}\right)  $ are the variances of the data points.

\subsection{Eisenhart-Duval lift}

From the reconstructed potential we construct the extended geometries of the
Riemannian and of the Lorentzian lift, and we calculate the Ricci scalar, the
Cotton-York tensor and the Weyl tensor by automatic differentiation of the
metric components. Automatic differentiation is exact up to machine precision.
Nevertheless, the curvature tensors depend on the second and third derivatives
of the learned potential $V\left(  q^{\mu}\right)  $, which are more sensitive
to the errors of the reconstruction than the potential itself.

\subsection{Random-feature network}

For the determination of the KVs and the CKVs of the extended metric we
introduce a neural network with random, non-trainable weights. Let $\xi^{A}$
be the generic CKV for the extended geometry expressed in a $\tanh$ base
\cite{hnn6}, such as
\begin{equation}
\xi^{A}\left(  \mathbf{X}\right)  =\sum_{j=1}^{N}a_{Aj}\,\phi_{j}\left(
\mathbf{X}\right)  ,\qquad\phi_{j}\left(  X\right)  =\tanh\left(  w_{j}%
\cdot\mathbf{X}+b_{j}\right)  ,\label{dd.06}%
\end{equation}
where $w_{j}$ are the weights and the biases $b_{j}$ are random numbers, and
only the coefficients $a_{Aj}$ are unknown coefficients. Parameter
$\mathbf{X}$ denotes the normalized coordinates $\left\{  q^{\mu},z\right\}  $
or $\left\{  q^{\mu},u,v\right\}  $ within the range $X\in\left[  -1,1\right]
$.

The symmetry condition $S\left(  a_{Aj}\right)  $ (\ref{sc.02}) is a linear
equation, which means that the derivation of the symmetry vectors is reduced
to a problem of optimizing the coefficients $a_{Aj}$. We evaluate the symmetry
condition at a large number of collocation points, and we define the least
square quantity, for each vector field $\xi_{k}$,
\begin{equation}
\sigma_{k}=\sqrt{\frac{S\left(  a_{Aj}\right)  ^{2}}{\xi_{k}^{2}}},~~\xi
_{k}^{2}=\xi_{k}^{A}\xi_{k~A},
\end{equation}
and we determine the coefficients $a_{Aj}$ which minimize it. We have
normalized to $\xi_{k}^{2}$ in order to avoid the determination of the trivial
symmetry $\xi_{k}^{A}=0$. A positive definite metric is necessary because for
the Lorentzian metric $\xi_{k}^{2}$ vanishes for the null vector fields, such
as the KV $\partial_{v}$.

For an exact symmetry it holds that $S\left(  a_{Aj}\right)  =0$, that is,
$\sigma_{exact}=0$. Nevertheless, for a learned metric the (trained)
symmetries provide small values for the parameter $\sigma$. For non-symmetry
vectors the obtained values $\sigma_{k}$ are far from zero, and when the
network finds non-symmetries, the evolution of the $\sigma_{k}$ has a gap. The
vector fields with $\sigma_{k}$ below the gap give the number of symmetries
and the corresponding KVs/CKVs.

Indeed, the network does not learn exact symmetries, but approximate
symmetries where $\sigma<\varepsilon$ and $\varepsilon$ is a small positive
number \cite{appr1}.

\subsection{Lie algebra and conservation laws}

From the learned symmetry vectors we can compute the commutators $\left[
\xi^{A},\xi^{B}\right]  $ by automatic differentiation. Because the learned
vector fields should form a Lie algebra, we calculate the structure constants
$C_{BC}^{A}$ by the least-square solution at the collocation points. This
allows us to test if the admitted symmetry vectors form a closed algebra.
\ Finally, we calculate the conservation laws by using Noether's theorem, that
is,
\begin{equation}
I\left(  \xi^{A}\right)  =\xi^{A}p_{A}\text{.}%
\end{equation}
Because the network has learned approximate symmetries, the corresponding
conserved quantities are approximate conservation laws \cite{appr1,appr2},
that is $\frac{dI}{dt}\simeq\Phi\left(  t\right)  $, with $\Phi^{2}\left(
t\right)  \rightarrow0$, a small varying function.

\section{ Experiments}

\label{sec4}

We proceed with the validation of the method described before by using mock
data generated from systems where the admitted symmetries and conservation
laws are known. Specifically, we consider the one-dimensional Ermakov-Pinney
equation described before, and the one-dimensional harmonic oscillator with or
without a quartic perturbation term.\ The last example allows us to explore
the application of the method to the construction of approximate conserved
quantities. For each of these systems we solve numerically the equations of
motion, and we add Gaussian noise on the position-momentum data points, with
different noise levels.

For this analysis we consider five independent networks and we report the
median obtained from the five different networks. For the Hamiltonian PINN we
consider a three layer network of sixty-four ($3\times64$) neurons each with a
$\tanh$ function, and the network was trained using the Adam optimizer. For
the random-feature network we used a sufficiently large number of collocation
points which depends on the dimension of the system, in order to recover the
complete number of symmetries.

\subsection{Ermakov 1D}

\begin{figure}[ptbh]
\centering\includegraphics[width=1\textwidth]{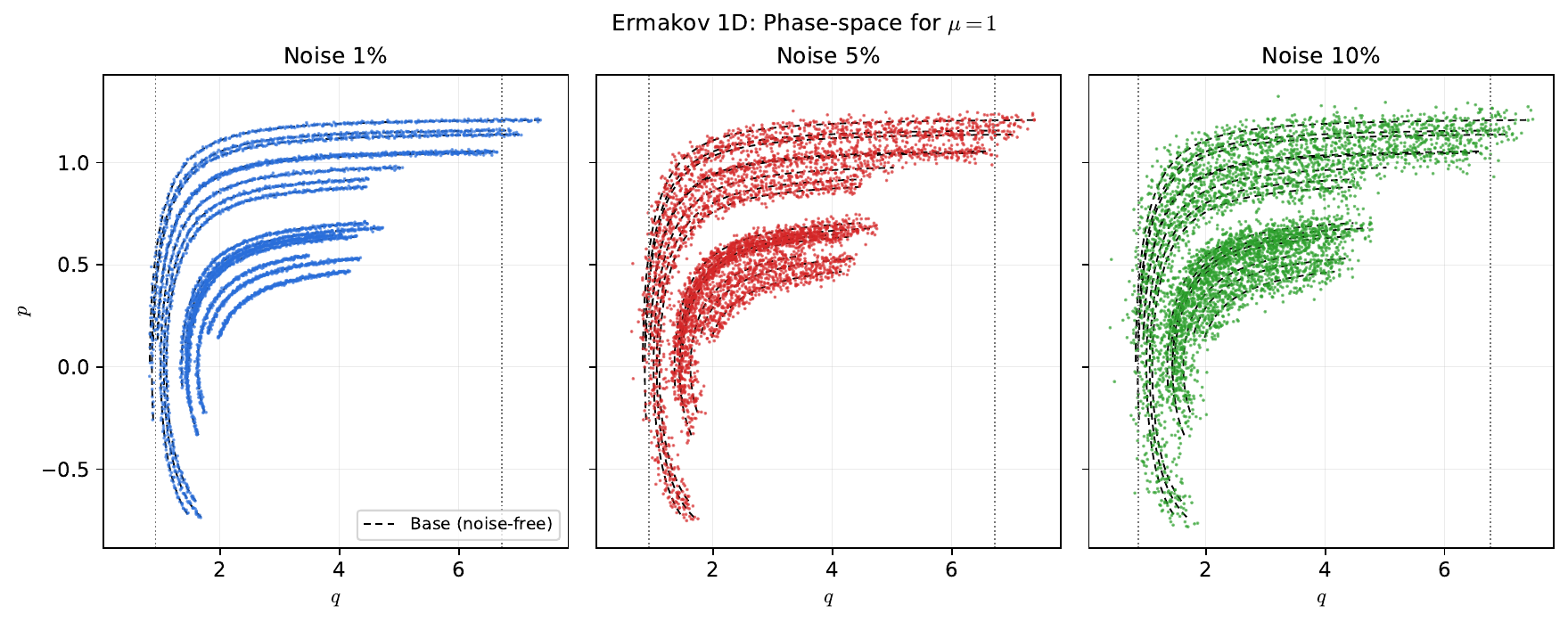}\caption{Ermakov
1D: Training data in the phase-space $\left(  q,p\right)  $, for the Gaussian
noise of $1\%,$ $5\%$ and $10\%$. }%
\label{fig1}%
\end{figure}

We consider the Ermakov-Pinney Hamiltonian system (\ref{ep.01}), with $\mu=1$.
We consider twenty trajectories with initial conditions $q_{0}\in\left[
0.5,2\right]  $ and $p_{0}\in\left[  -0.8,0.2\right]  $ and we integrated so
as to create 300 points per trajectory. Then, we introduced Gaussian noise to
the data $\left(  q\left(  t_{k}\right)  ,p\left(  t_{k}\right)  \right)  $.
Specifically, we have added three different levels of noise, of $1\%$, $5\%$
and $10\%$, on the same baseline trajectories, as can be seen in Fig.
\ref{fig1}.

\begin{figure}[ptbh]
\centering\includegraphics[width=1\textwidth]{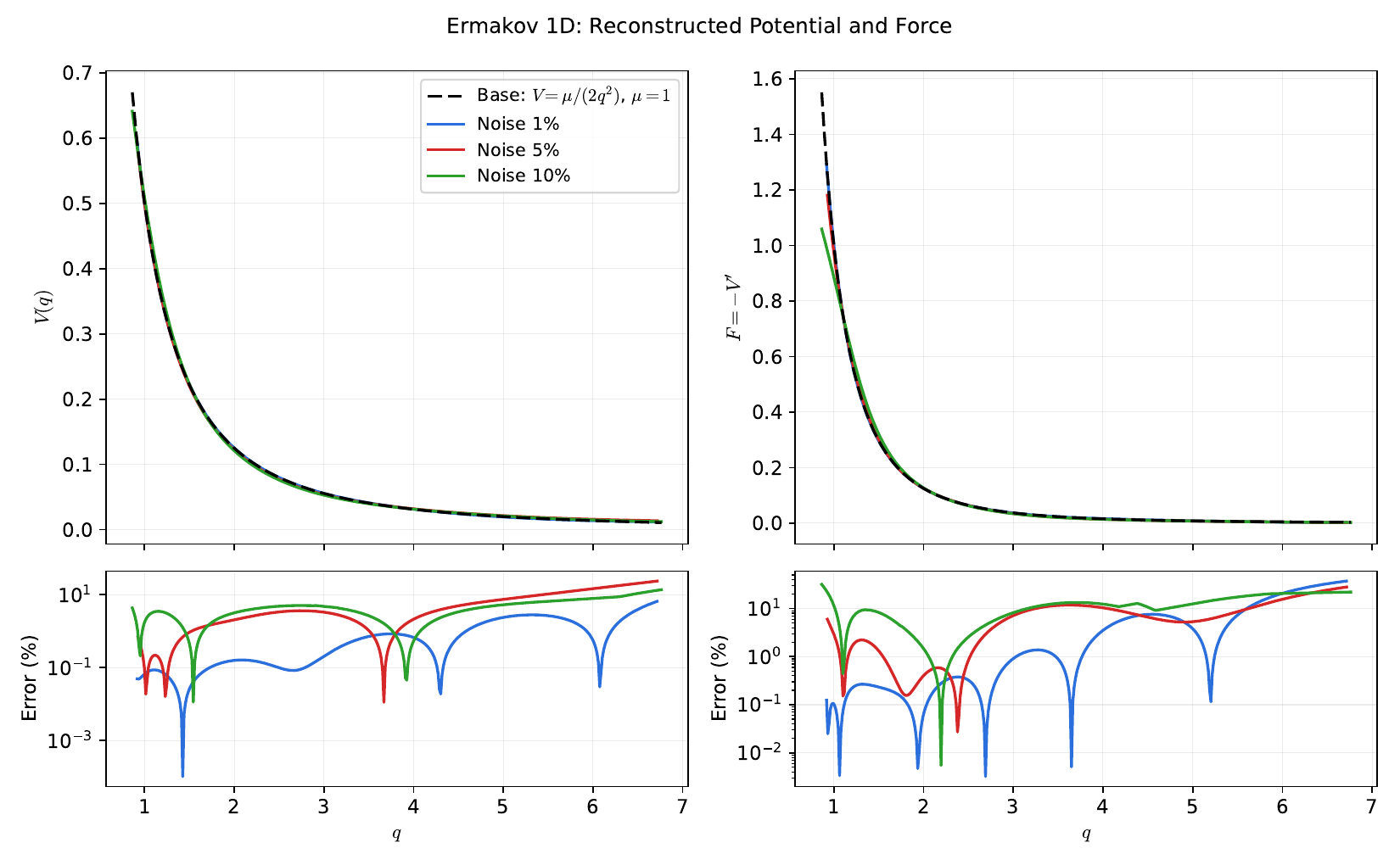}\caption{Ermakov
1D: Reconstructed potential function and force term from the Hamiltonian PINN
for noise levels $1\%$, $5\%$ and $10\%$. The lower panels give the relative
error.}%
\label{fig2}%
\end{figure}

In Fig. \ref{fig2} we present the reconstruction of the potential function,
and of the force term, as follows from the training of the Hamiltonian PINN.
We observe that over the region covered by the data, the typical error is
lower than the noise level. The noise in the force term arises from the
automatic differentiation used to compute it. Note that the potential is the
function being trained. 

\begin{figure}[ptbh]
\centering\includegraphics[width=0.5\textwidth]{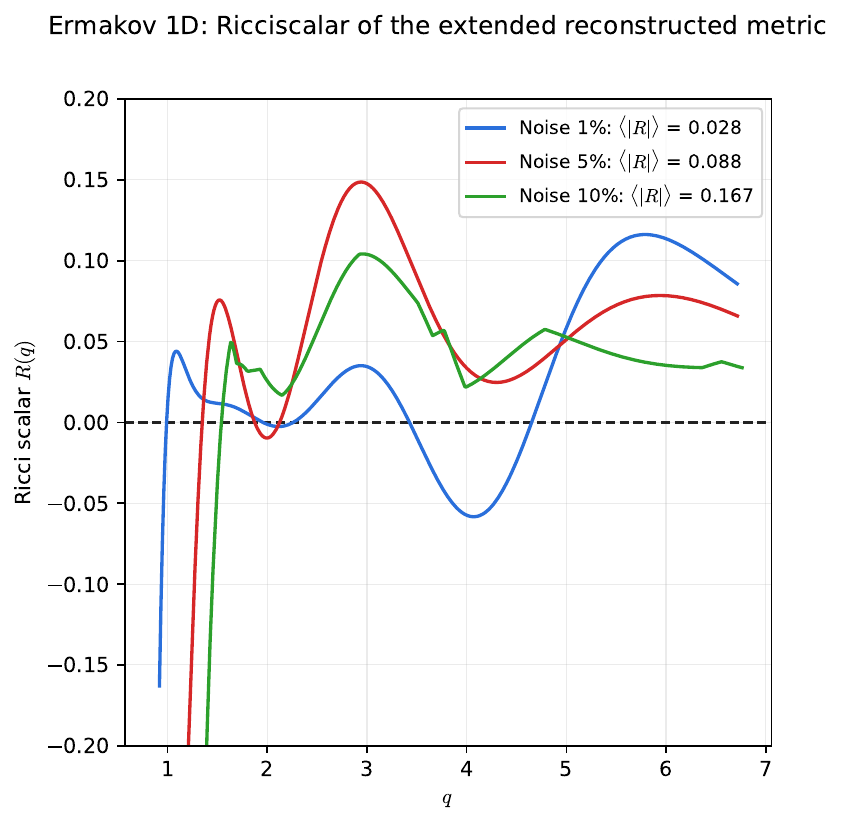}\caption{Ermakov
1D: Derived Ricci scalar $R\left(  q\right)  $ for the reconstructed extened
line elemenet after the application of the Riemannian lift. }%
\label{fig3}%
\end{figure}

We employ the Riemannian lift, and for the reconstructed two-dimensional
extended metric with line element (\ref{ns.04}) we calculate the Ricci scalar
$R\left(  q\right)  $ as presented in Fig. \ref{fig3}. The Ricci scalar is
determined by the second-derivative of the potential $V\left(  q\right)  $,
where the automatic differentiation is employed. For $1\%$ and $5\%$ noise,
the Ricci scalar stays close to the zero value with averaged values
\thinspace$\left\vert R\right\vert =0.028$ and $\left\vert R\right\vert
=0.088$ for the $1\%$ and $5\%$ noise levels respectively. On the other hand,
for the data with $10\%$ noise, the reconstructed Ricci scalar had averaged
value $\left\vert R\right\vert =0.167$, which indicates a deviation from the
zero value, as expected for the analytic form. The main deviation exists for
$q<1$, as the trajectories reach the center. We conclude that the
reconstructed extended metric is thus flat to within the reconstruction error.
This should also be verified by the number of admitted KVs, which should be
three and form the $e\left(  2\right)  $ Lie algebra. 

\begin{figure}[ptbh]
\centering\includegraphics[width=0.75\textwidth]{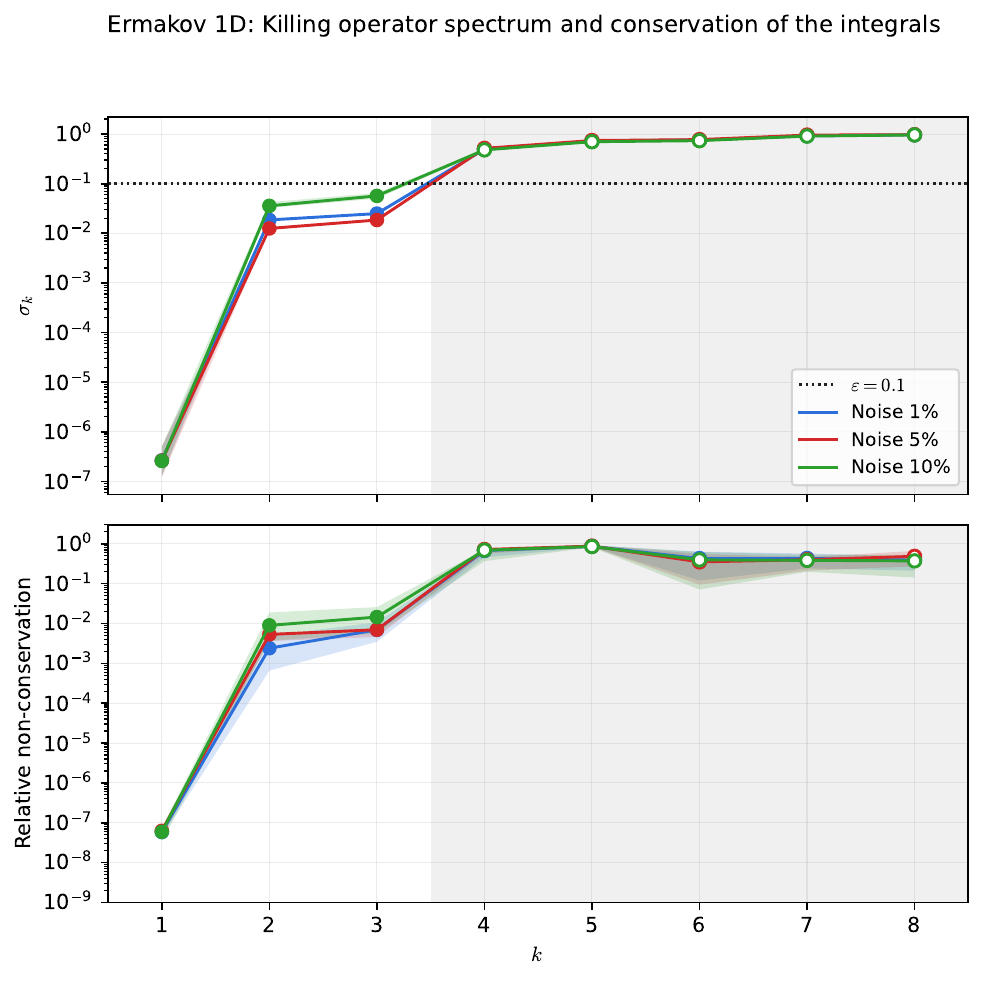}\caption{Ermakov
1D: Values $\sigma_{k}$ of the Killing operator of the reconstructed extended
metric. In the lower-pane we present the relative non-conservation $r_{k}$ of
the quantity $I_{k}=\xi_{k}^{A}p_{A}$ on different trajectores. }%
\label{fig4}%
\end{figure}

Within the reconstructed extended geometry we apply the random-feature network
for the derivation of the KVs. In Fig. \ref{fig4} we present the lowest values
of the $\sigma$, for three noise levels. The markers are the median of the
twenty independent trajectories. For each noise level we observe a similar
behaviour. There are three vector fields with $\sigma<0.1$, while for $k>3$,
the $\sigma\simeq1$, from where we can conclude that the network has
reconstructed the three KVs of the two-dimensional flat space. The lower-panel
of \ref{fig4} shows the relative non-conservation of the quantity $r_{k}%
=\frac{d}{dt}\ln\left(  I_{k}\left(  \xi_{k}^{A}\right)  \right)  $ , where we
see that for $k=1,2$ and $3$, the values are smaller than $10^{-1}$, while for
$k>3$, the non-conservation value is close to $10^{0}$. \ 

\begin{figure}[ptbh]
\centering\includegraphics[width=1\textwidth]{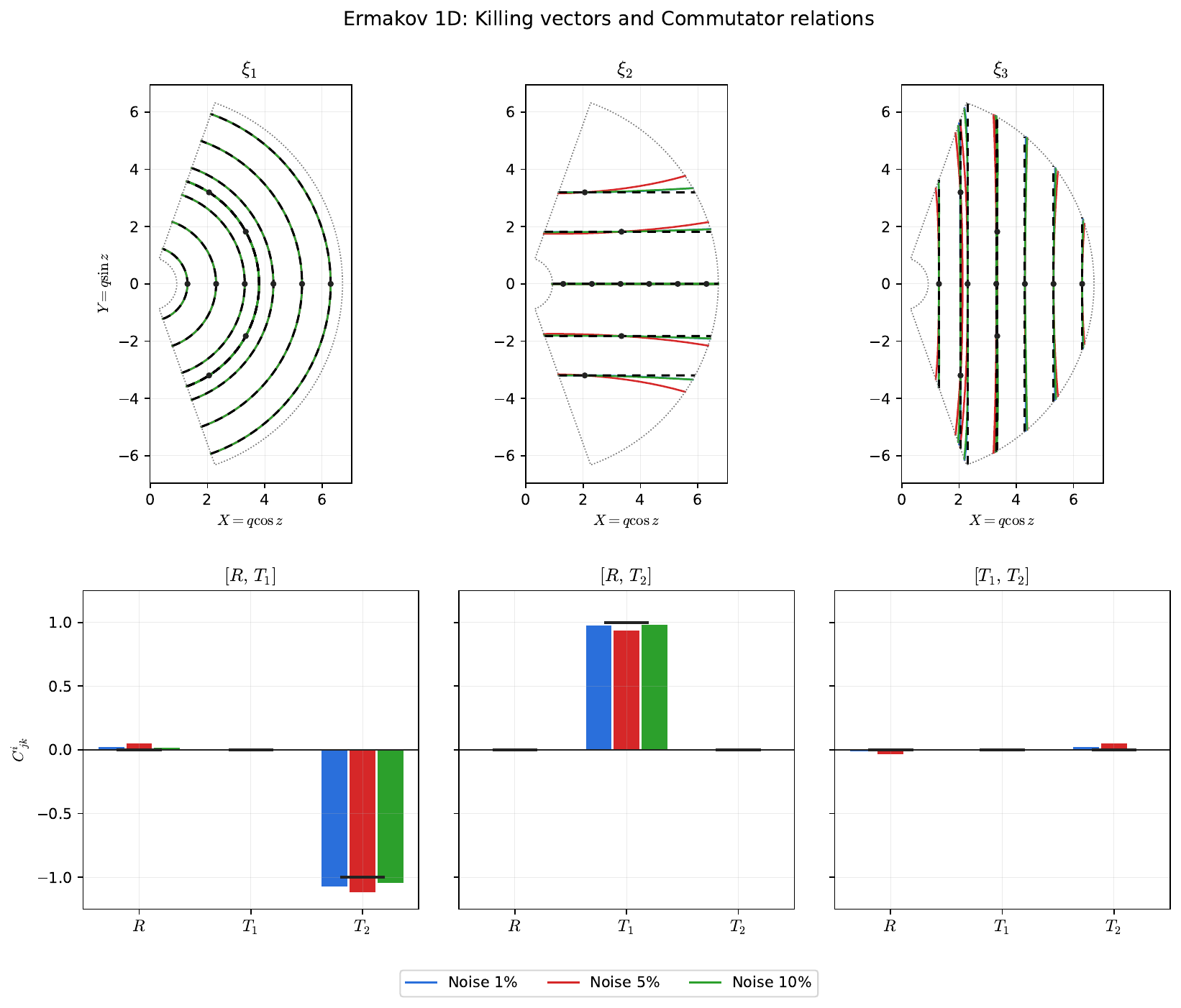}\caption{Ermakov
1D: Integral curves of the three learned KVs and commutator relations.}%
\label{fig5}%
\end{figure}

\begin{figure}[ptbh]
\centering\includegraphics[width=1\textwidth]{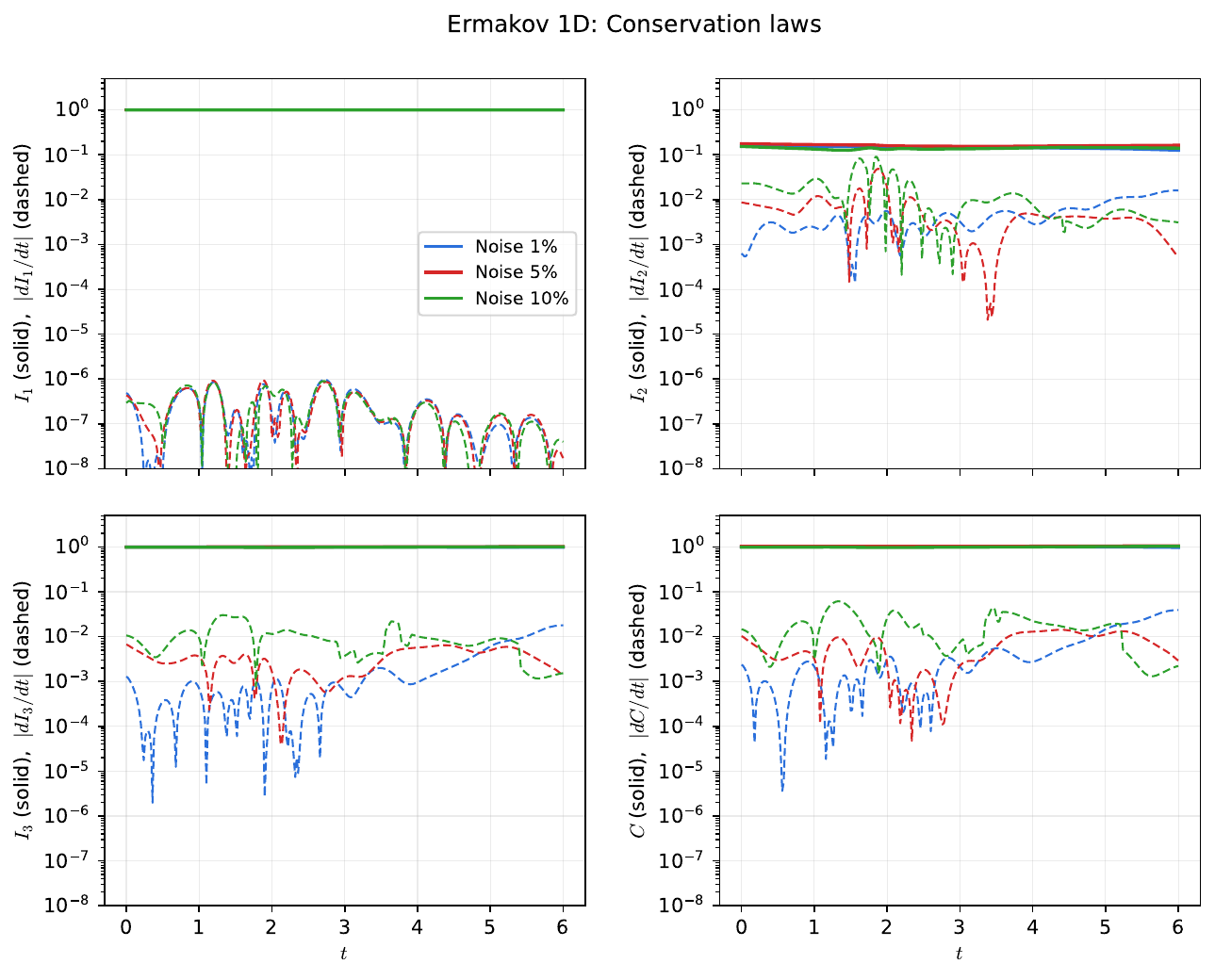}\caption{Ermakov
1D: Conservation laws for the geodesic equations and Casimir invariant. }%
\label{fig7}%
\end{figure}

The symmetry which corresponds to $k=1$ is the trivial conservation law of the
momentum $p_{z}$, which is the angular momentum and corresponds to the
rotation symmetry.\ The other two vector fields correspond to the
two-translations of the flat space. In Fig. \ref{fig5} we plot the\ interval
curves for the reconstructed KVs in Cartesian coordinates, and we present the
commutators of the obtained vector fields. Finally, in \ref{fig7} we present
the derivation of the conservation law and its evolution $\frac{dI}{dt}$ for a
specific trajectory, as well as the Casimir invariant of the admitted Lie algebra.

We conclude that the reconstructed space is (approximately) flat.

\subsection{Harmonic Oscillator 1D}

\begin{figure}[ptbh]
\centering\includegraphics[width=1\textwidth]{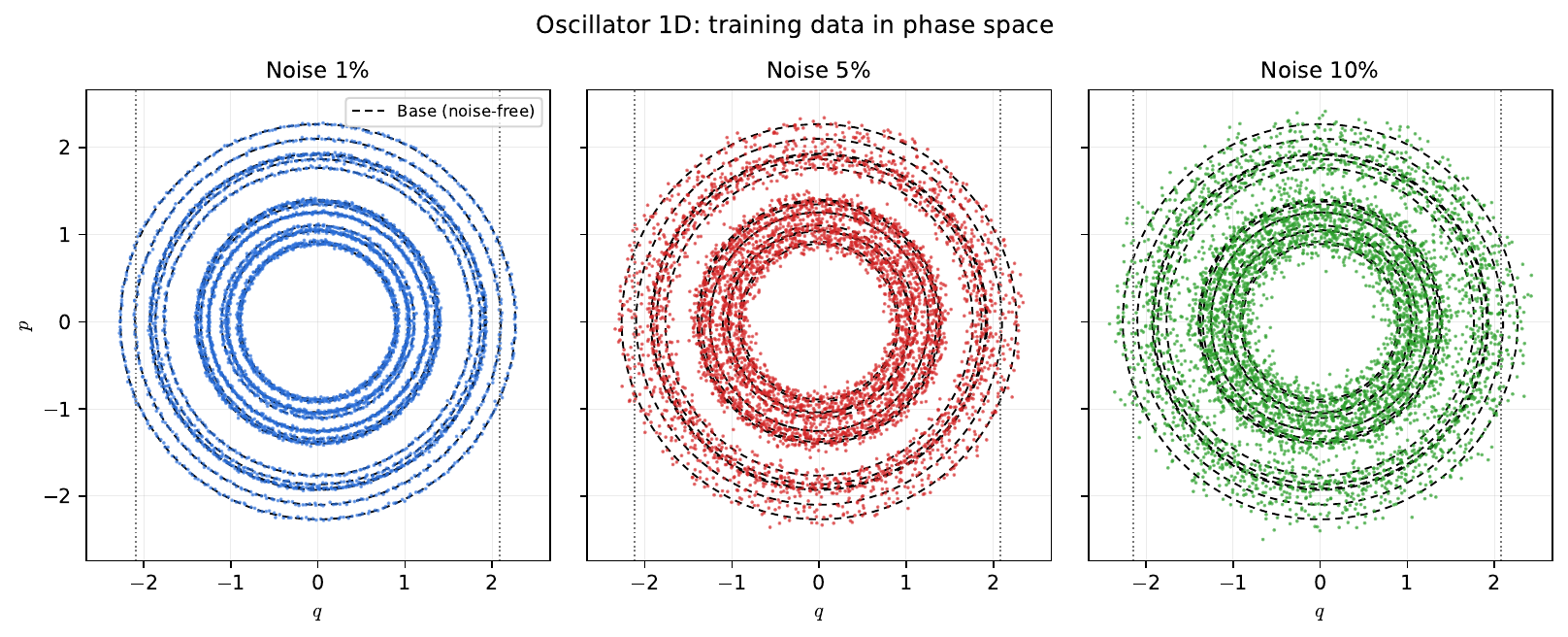}\caption{Harmonic
Oscillator 1D: Training data in the phase-space $\left(  q,p\right)  $, for
the Gaussian noise of $1\%,$ $5\%$ and $10\%$. }%
\label{fig9}%
\end{figure}

We proceed with the derivation of the conservation laws for the
one-dimensional harmonic oscillator. We consider again data from twenty
trajectories as presented in the phase-space portraits of Fig. \ref{fig9}. The
initial conditions of the trajectories are for $q_{0}\in\left[  -2,2\right]
$, $~p_{0}\in\lbrack-2,2]$ and $\omega^{2}=1.$ The data are for trajectories
of one period.

\begin{figure}[ptbh]
\centering\includegraphics[width=1\textwidth]{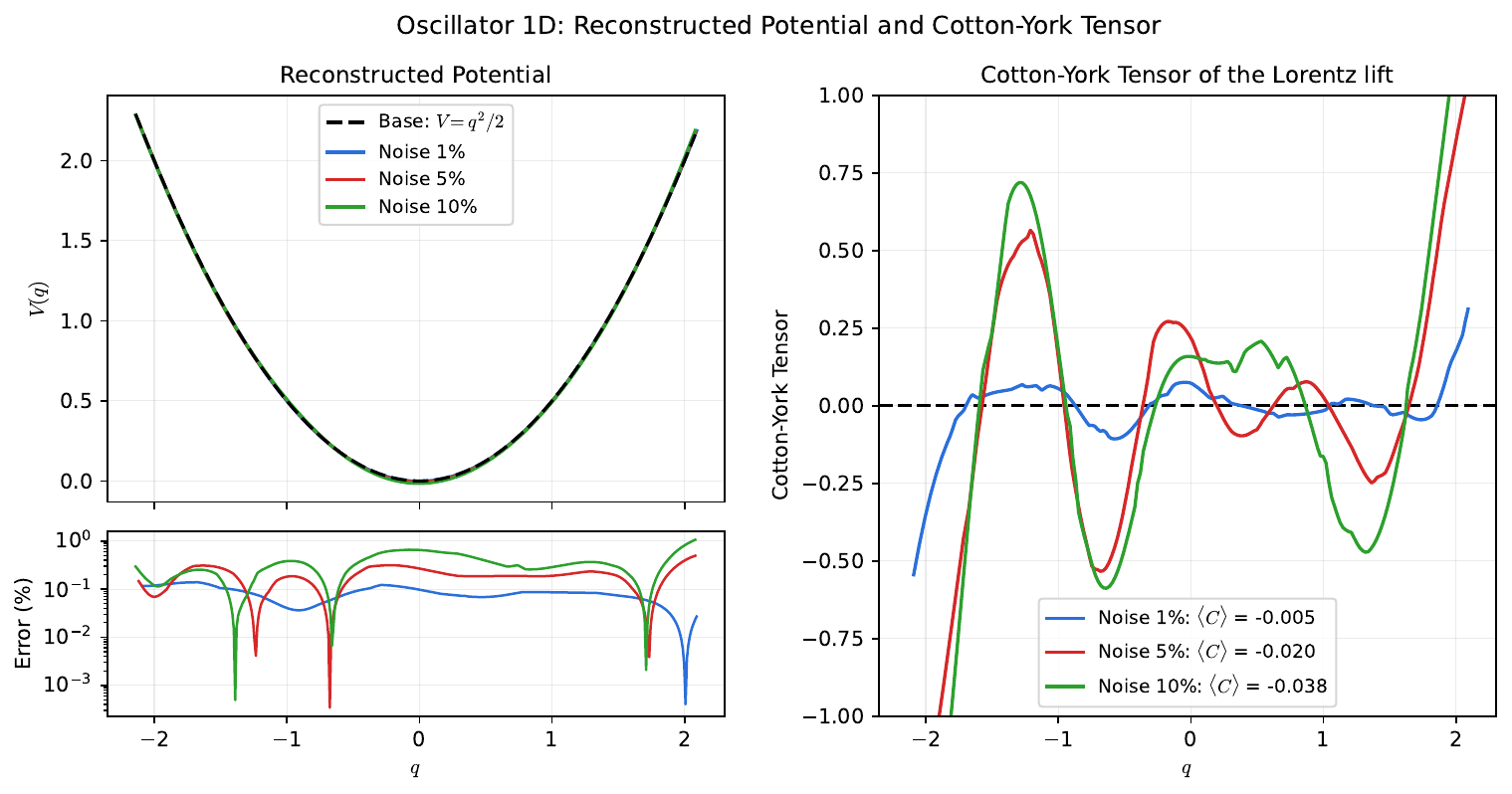}\caption{Harmonic
Oscillator 1D: Reconstructed potential function from the Hamiltonian PINN for
noise levels $1\%$, $5\%$ and $10\%$. The lower panel gives the relative
error. Moreover, we present the reconstructed Cotton-York tensor for the
Lorentzian extended metric, which is derived with automatic differentiation of
the reconstructed potential. }%
\label{fig10}%
\end{figure}

Using the Hamiltonian PINN, we reconstruct the potential function
and compute the Cotton--York tensor of the extended geometry (\ref{nn.11})
obtained from the Lorentzian lift. The latter are shown in Fig. \ref{fig10},
from which we see that the network reconstructs the potential function well.
Although the Cotton-York tensor is not zero, it has an average value
$\left\langle C\right\rangle $ close to zero with values
$-0.005,-0.020,-0.038$ for the $1\%$, $5\%$ and $10\%$ noise levels
respectively. Hence, the reconstructed extended geometry is asymptotically
conformally flat for all the noise levels. The amplitude of the oscillations
for the Cotton-York tensor depends on the noise level. This reflects the
sensitivity of the Cotton-York tensor to small errors in the learned
potential, since the Cotton-York tensor follows from the third derivative of
the potential, and repeated differentiation amplifies small deviations of the
network from the exact potential.

\begin{figure}[ptbh]
\centering\includegraphics[width=0.75\textwidth]{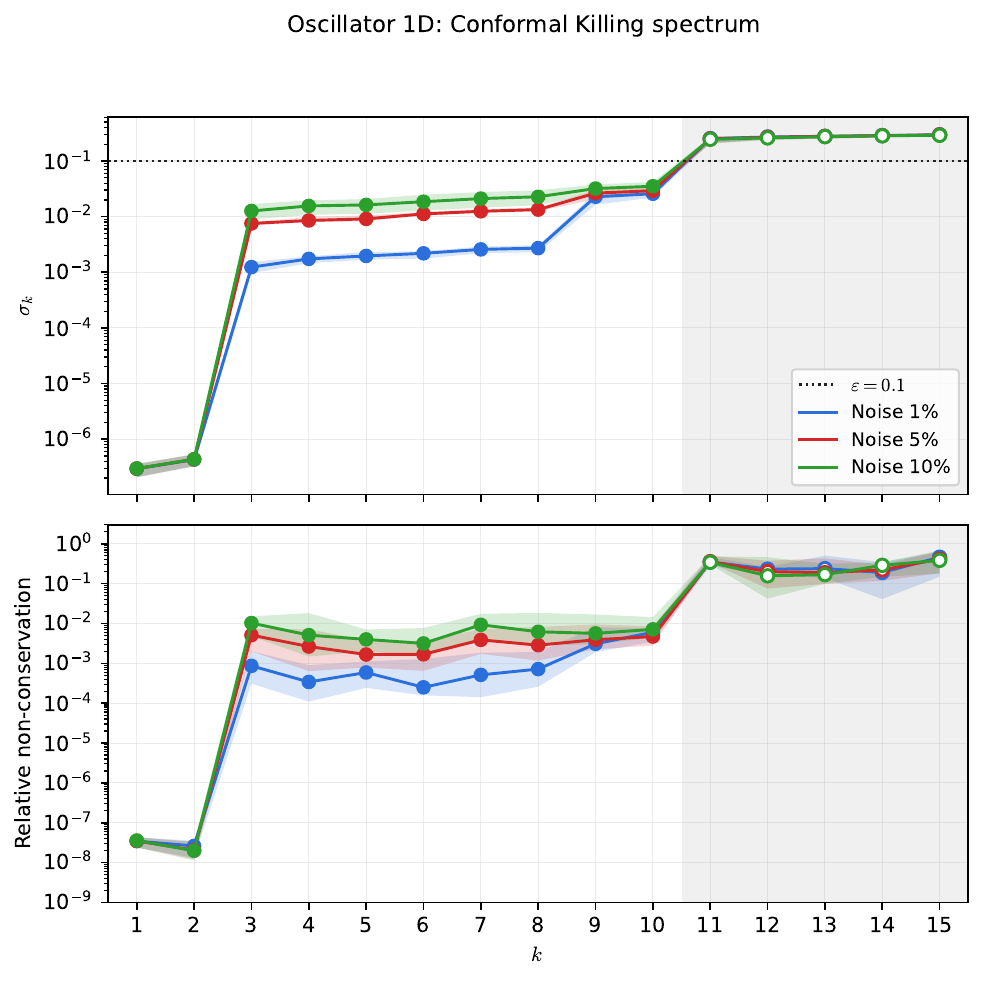}\caption{Harmonic
Oscillator 1D: Values $\sigma_{k}$ of the Conformal Killing operator of the
reconstructed extended metric. In the lower-pane we present the relative
non-conservation $r_{k}$ of the quantity $I_{k}=\xi_{k}^{A}p_{A}$ on different
trajectores. }%
\label{fig11}%
\end{figure}

\begin{figure}[ptbh]
\centering\includegraphics[width=1\textwidth]{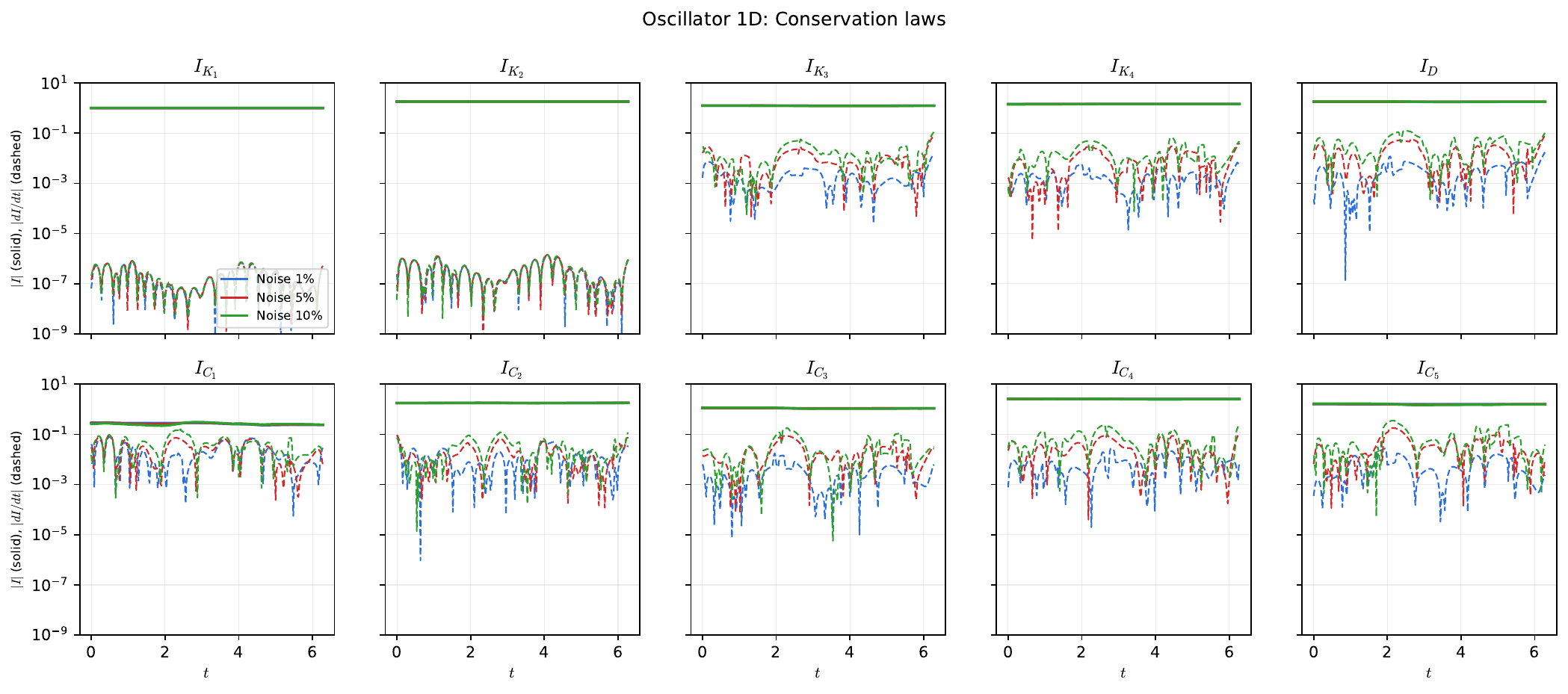}\caption{Harmonic
Oscillator 1D: Conservation laws for the geodesic equations.}%
\label{fig12}%
\end{figure}

\begin{figure}[ptbh]
\centering\includegraphics[width=0.75\textwidth]{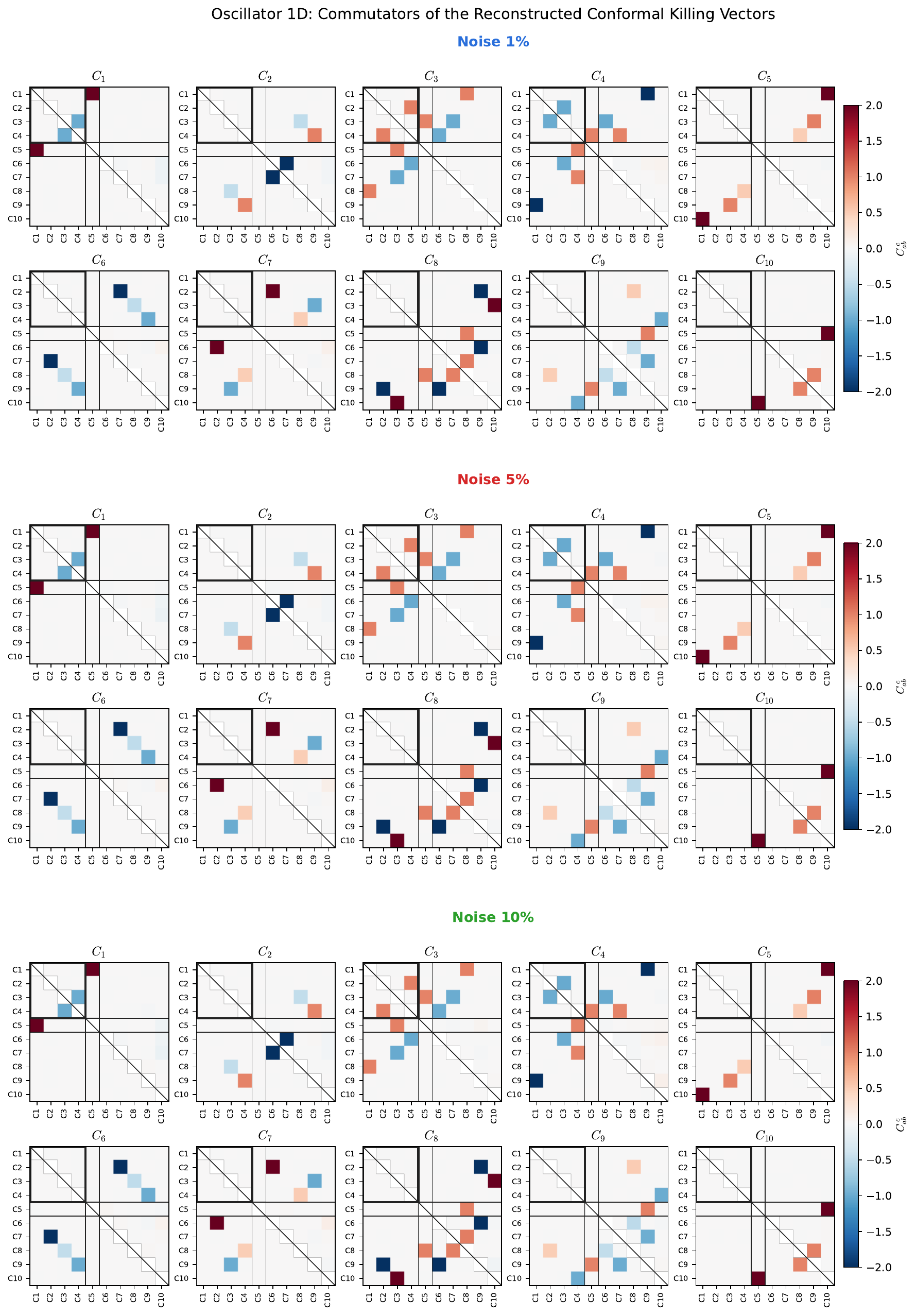}\caption{Harmonic
Oscillator 1D: Commutators for the reconstructed ten CKVs. The vector fields
form the $so\left(  3,2\right)  $ Lie algebra, independent of the noise level.
}%
\label{fig13}%
\end{figure}

In Fig. \ref{fig11} we present the values for the $\sigma$ as derived for the
CKV condition, for the reconstructed three-dimensional extended metric.
Furthermore, we present the non-conservation $r_{k}=\left(  rms_{t}\left(
I_{k}\left(  \xi_{k}^{A}\right)  \right)  \right)  ^{-1}\frac{d}{dt}%
I_{k}\left(  \xi_{k}^{A}\right)  $ of the quantities $I_{k}=\xi_{k}^{A}p_{A}$.
We observe that for $k\leq10$, the $\sigma_{k}\leq10^{-1}$, while for $k>10$,
$\sigma_{k}$ are close to $10^{0}$, while for the $r_{k}$ it follows that for
$k\leq10$,~$r_{k}<10^{-2}$, while for $k>10$ the quantities $I_{k}$ are not
conservation laws. The time evolution for the ten admitted conserved
quantities is plotted for trained trajectories in Fig. \ref{fig12}. Moreover,
Fig. \ref{fig13} shows the commutators for the reconstructed CKVs where we see
that the commutator relations are independent of the noise level, and they
form the $so\left(  3,2\right)  $ Lie algebra, as expected for the
three-dimensional conformally flat geometry.

\subsubsection{Quartic correction}

We introduce a quartic term $V_{corr}\left(  q\right)  =\epsilon q^{4}$ to the
one-dimensional harmonic oscillator and repeat the analysis. We assume two
cases, $\epsilon=10^{-3}$ and $\epsilon=10^{-1}$. For $\epsilon\neq0$, the
dynamical system does not admit any nontrivial symmetries, except for the
time-translation symmetry with conservation law the Hamiltonian function.
Consequently, the extended three-dimensional metric of the Lorentzian lift has
only two CKVs, the vector fields $\partial_{u}$ and $\partial_{v}$. The other
eight symmetries of the oscillator are broken by the correction term and
survive only as approximate symmetries, to a degree controlled by $\epsilon$.

\begin{figure}[ptbh]
\centering\includegraphics[width=0.75\textwidth]{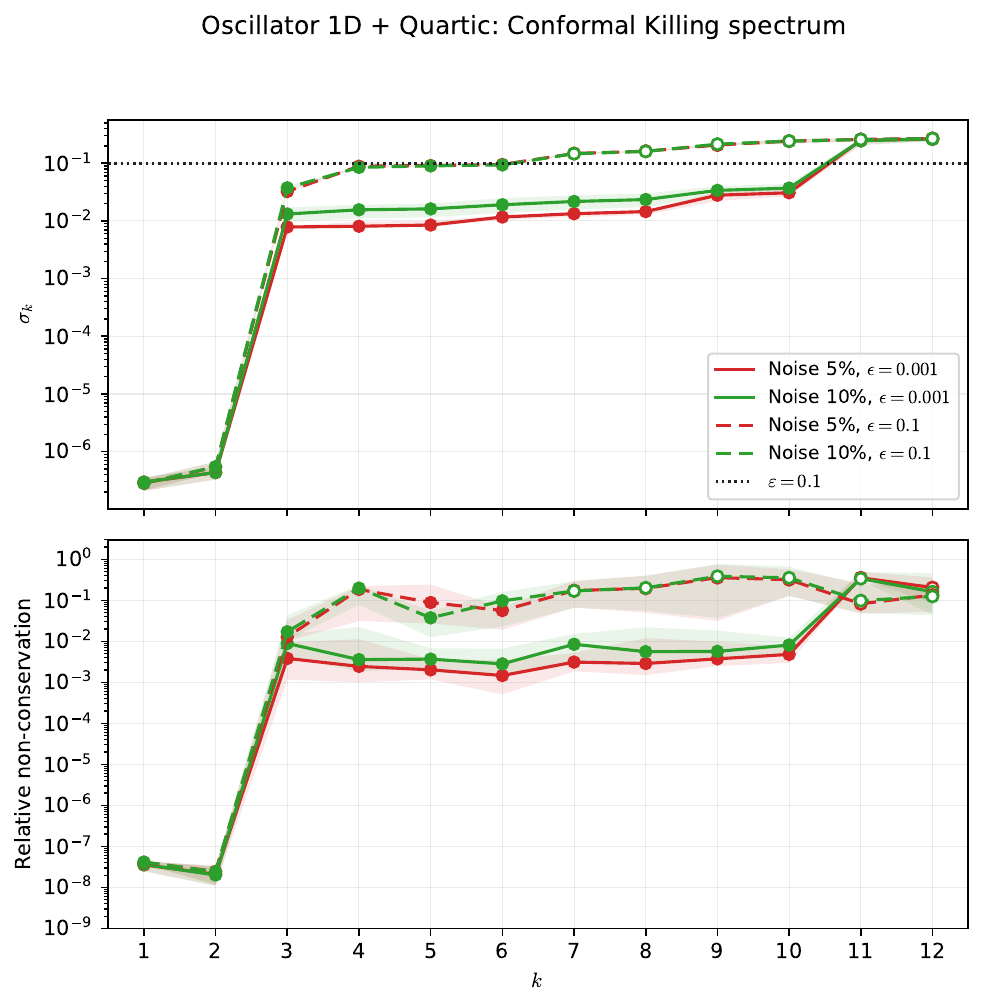}\caption{Harmonic
Oscillator 1D with Quartic Correction: Values $\sigma_{k}$ of the Conformal
Killing operator of the reconstructed extended metric. In the lower-pane we
present the relative non-conservation $r_{k}$ of the quantity $I_{k}=\xi
_{k}^{A}p_{A}$ on different trajectores. }%
\label{fig14}%
\end{figure}

In Fig. \ref{fig14} we present the values for the $\sigma$ as derived for the
CKV condition, for the reconstructed three-dimensional extended metric, and
the relative non-conservation $r_{k}$ for the corresponding quantities along
the (null) geodesic equations. For the small correction term $\epsilon
=10^{-3}$ and for the time evolution of one period, we see that there are ten
vector fields where $\sigma<10^{-1}$, and the relative non-conservation
$r_{k}$ are lower than $10^{-2}$. On the other hand, for $\epsilon=10^{-1}$,
there are only six vector fields with $\sigma<10^{-1}$, where the
corresponding relative non-conservation $r_{k}$ are larger than $10^{-2}$. For
the $\epsilon=10^{-3}$ we observe the gap in the $\sigma$, for $k>10$; such a
gap does not exist for the $\epsilon=10^{-1}$ case.

\section{Conclusions}

\label{sec5}

In this work we introduced a data-driven framework for the derivation of
symmetries and of conservation laws that are linear in the momenta, for
Newtonian dynamical systems described by Hamiltonian functions with kinetic
and potential terms. We employed the Eisenhart-Duval lift to geometrize the
dynamical system, which reduces the derivation of the symmetries to the
determination of the collineations of an extended geometry, and specifically
to the determination of its Killing and conformal Killing algebras.

Our approach consists of four steps. First, a Hamiltonian PINN learns the
potential function from phase-space data. Second, the learned potential
defines the extended metric of the Riemannian or of the Lorentzian lift, and
we calculate the curvature invariants of these geometries, which provide
important information about the structure of the geometry and the nature of
the geodesic equations. Third, a random-feature network solves the Killing and
the conformal Killing equations. Finally, the Lie algebra and the Noetherian
conservation laws are determined from the learned vector fields.

The main advantage of the geometric formulation is that the symmetry
conditions for the original equations of motion, in which the potential enters
through additional constraints, are replaced by the Killing and conformal
Killing equations of the extended metric. These equations are linear in the
unknown vector fields, so that the symmetry problem becomes a linear algebra
problem and the number of symmetries can be read from the spectrum of the
operator. Moreover, the conservation laws which follow from the extended
geometry are not necessarily local functions of the original variables, which
means that the approach also reveals nonlocal conservation laws.

In order to demonstrate the application of the framework, we considered mock
data for the one-dimensional Ermakov-Pinney equation and the one-dimensional
harmonic oscillator. The two examples were not selected randomly. Through the
Eisenhart-Duval lift, both lead to geodesic equations in geometries with the
maximal number of symmetries. This allows us to test our framework in the
reconstruction of the symmetries of geometries with a maximal symmetry
algebra. Specifically, the Ermakov-Pinney equation is equivalent to the
geodesic equations of the two-dimensional flat space, which admits the maximal
Killing algebra, while the harmonic oscillator is equivalent to the null
geodesic equations of a three-dimensional conformally flat Lorentzian
geometry, which admits the maximal conformal algebra.

For different levels of Gaussian noise on the solution trajectories of the
aforementioned systems, we reconstructed the potential function and the
geometric features of the extended geometries up to a certain noise level.
Furthermore, we showed that the random-feature neural network reconstructs the
full symmetry algebra, of maximal dimension, admitted by the equivalent
geodesic equations.

The present analysis has several limitations. First, we assumed a specific
form of the Hamiltonian function, in which the kinetic metric is flat and
expressed in Cartesian coordinates. Second, we focused only on conservation
laws which are linear in the momenta and are constructed from the KVs and the
CKVs, and we considered only the Riemannian and the Lorentzian lifts, although
other lifts exist which can provide new results \cite{annew}. Moreover, the
random-feature neural network learns the symmetries locally, so the
reconstructed conservation laws are valid only within the region covered by
the data.

\begin{figure}[ptbh]
\centering\includegraphics[width=0.75\textwidth]{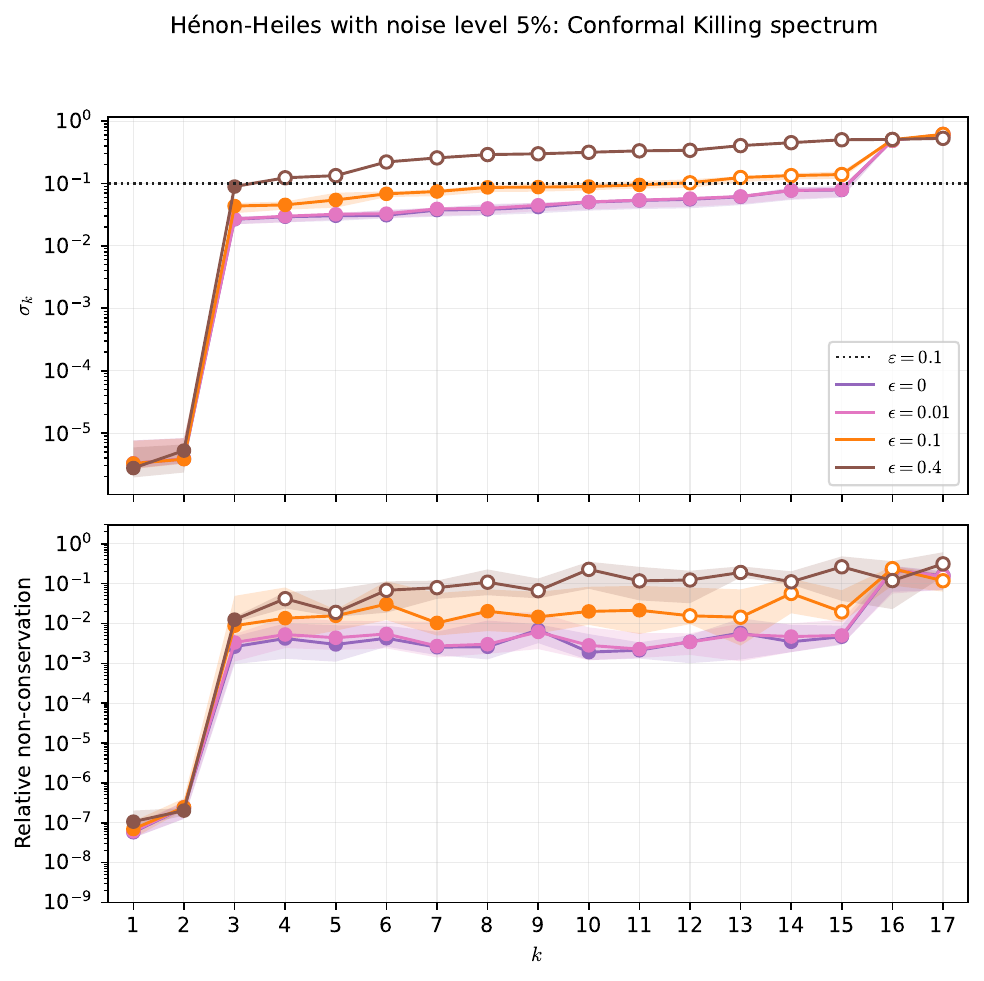}\caption{H\'{e}non-Heiles:
Values $\sigma_{k}$ of the Conformal Killing operator of the reconstructed
extended metric. In the lower-pane we present the relative non-conservation
$r_{k}$ of the quantity $I_{k}=\xi_{k}^{A}p_{A}$ on different trajectores. }%
\label{fig15}%
\end{figure}

In order to demonstrate the application of the framework to higher-dimensional
systems, we considered the two-dimensional H\'{e}non-Heiles system, as a
perturbation of the two-dimensional isotropic harmonic oscillator by the
correction term $V_{corr}^{HH}=\epsilon\left(  q_{1}^{2}q_{2}-\frac{q_{2}^{3}%
}{3}\right)  $. We applied the framework for $\epsilon=0$, $10^{-2}$,
$10^{-1}$ and $4\times10^{-1}$, with $5\%$ noise, for trajectories for half a
period of the oscillator. For $\epsilon=0$ the extended Lorentzian metric is
four-dimensional and conformally flat, and it admits the maximal,
fifteen-dimensional, conformal algebra $so(4,2)$. Fig. \ref{fig15} shows the
spectrum $\sigma_{k}$ of the conformal Killing operator and the relative
non-conservation $r_{k}$. For $\epsilon=0$ we recover the maximal number of
symmetries, fifteen vector fields have $\sigma_{k}<10^{-1}$, and the spectrum
has a clear gap at $k=16$. For $\epsilon=10^{-2}$ we recover the same number
of symmetries, with values of $r_{k}$ comparable to those of the $\epsilon=0$
case. For $\epsilon=10^{-1}$ a weaker gap remains at $k=16$, while the values
of $r_{k}$ increase to approximately $10^{-2}$. Finally, for $\epsilon
=4\times10^{-1}$ the gap disappears, and only the two exact symmetries remain,
which correspond to the trivial vector fields $\partial_{u}$ and $\partial
_{v}$.

In a future work we plan to extend this framework to the reconstruction of
higher-order conservation laws and to more general forms of the Hamiltonian
function. Conservation laws which are quadratic or higher-polynomiam functions
in the momenta, such as in the the integrable cases of the H\'{e}non-Heiles
system, the Lewis-invariant for the higher-dimensional Ermakov-Pinney system,
or the Runge-Lenz vector of the Kepler problem, are a result of Killing
tensors of the extended metric or conformal Killing tensors for the case of
null geodesics.

Finally, the geometric formulation offers a natural route to quantization. The
symmetries learned from the classical trajectories can be used for the
construction of the symmetry operators and of the conserved quantities of the
quantum system, and for the study of the relation between classical and
quantum integrability directly from data.

\bigskip\textbf{Code Availability Statement:} The code is
available upon reasonable request.

\bigskip\textbf{Declaration of competing interests:}The author declare no conflict of interests.

\begin{acknowledgments}
AP was partially supported by FONDECYT Grant No. 1240514.
\end{acknowledgments}

\end{document}